 \documentclass[aps,prb,twocolumn,groupedaddress]{revtex4}

\usepackage{graphicx}
\usepackage{ams}

\begin{document}

\title{\textbf{Comparison of chain versus sheet crystal structures for  \\
cyanides $M$CN ($M$ = Cu-Au) and dicarbides $M$C$_2$ \\
($M$ = Be-Ba; Zn-Hg). Alternatives to graphene ?}}

\author{P.~Zaleski-Ejgierd\thanks{E-mail address:
Patryk.Zaleski-Ejgierd@helsinki.fi}}\affiliation{ Laboratory for
Instruction in Swedish, Department of Chemistry, P.O.B. 55,
FIN-00014 University of Helsinki, Finland}

\author{M.~Hakala\thanks{E-mail address:
Mikko.O.Hakala@helsinki.fi}}\affiliation{Division of X-ray Physics,
Department of Physical Sciences, P.O.B. 64, FIN-00014 University of
Helsinki, Finland}

\author{P.~Pyykk\"o\thanks{E-mail address:
Pekka.Pyykko@helsinki.fi}}\footnote{To whom correspondence should be
addressed. E-mail: Pekka.Pyykko@helsinki.fi}\affiliation{ Laboratory
for Instruction in Swedish, Department of Chemistry, P.O.B. 55,
FIN-00014 University of Helsinki, Finland}

\date{\today}

\begin{abstract}
The cyanides $M$CN, $M$=Cu, Ag, Au, have experimentally a structure
with hexagonally packed, infinite -$M$-CN-$M$-CN- chains. Following
our earlier study for AuCN, we now predict that all three $M$CN could
have an alternative $M_3$C$_3$N$_3$ sheet structure of comparable
energy with the known one. The valence isoelectronic systems $M$C$_2$
versus $M_3$C$_6$, $M$=Be-Ba; Zn-Hg are also studied. Now, the known
dicarbides have the CaC$_2$ or MgC$_2$ chain structures. The predicted
sheets lie energetically below the chains for $M$ = Zn, Cd, and
Hg. All these systems are experimentally unknown. Indeed, they are
clearly endothermic, compared to the elements. For some sheet
structures the densities of states suggests rather small band gaps and
even metallic character. When available, the experimental geometries
agree well with the calculated ones for both cyanides and dicarbides.

\end{abstract}



\maketitle

\section{Introduction}
\label{sec:introduction} 
\begin{figure}[h]
\includegraphics[width=5.5cm,angle=-90,scale=0.70]{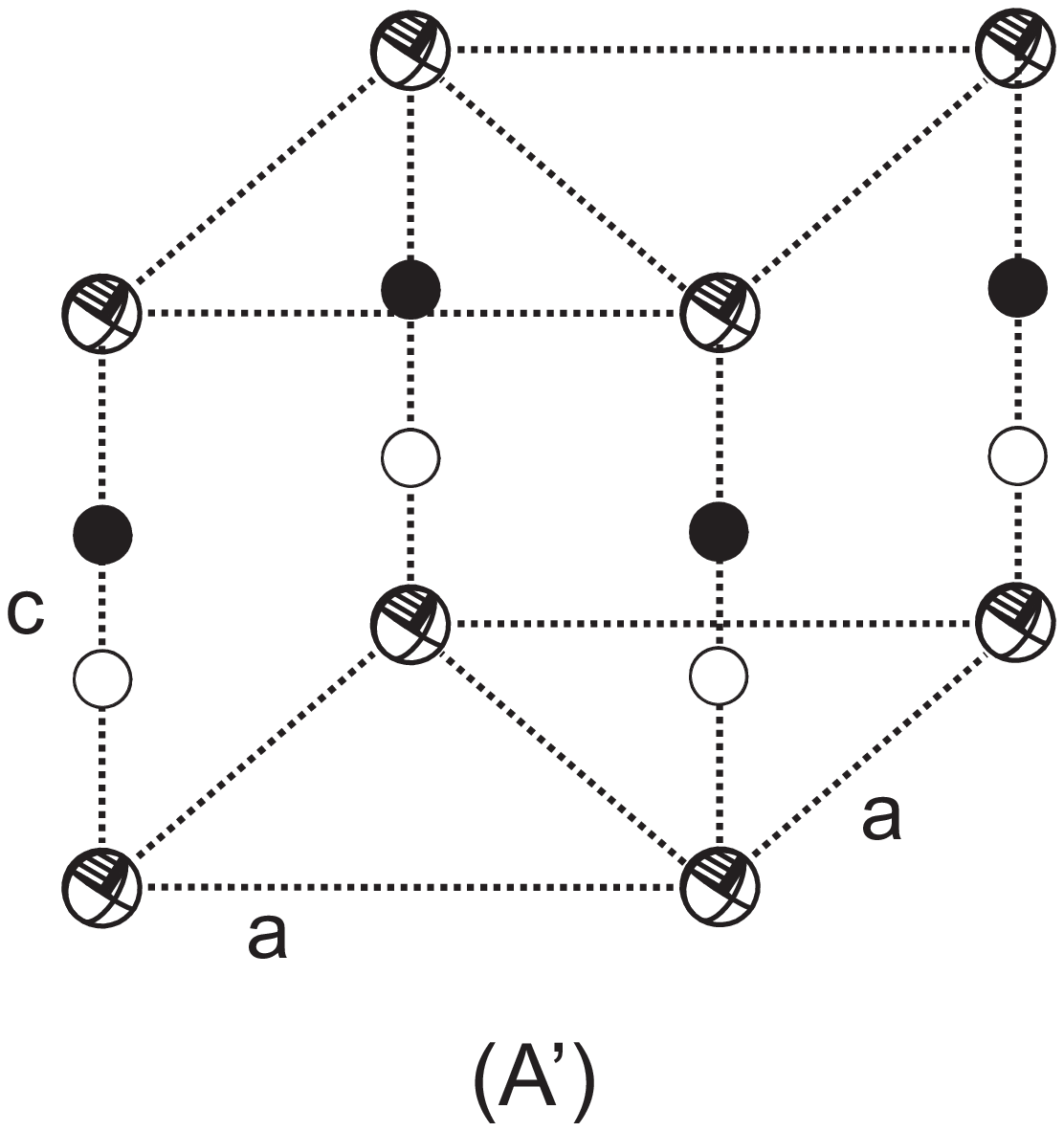}\hspace{3mm}
\includegraphics[width=5.5cm,angle=-90,scale=0.70]{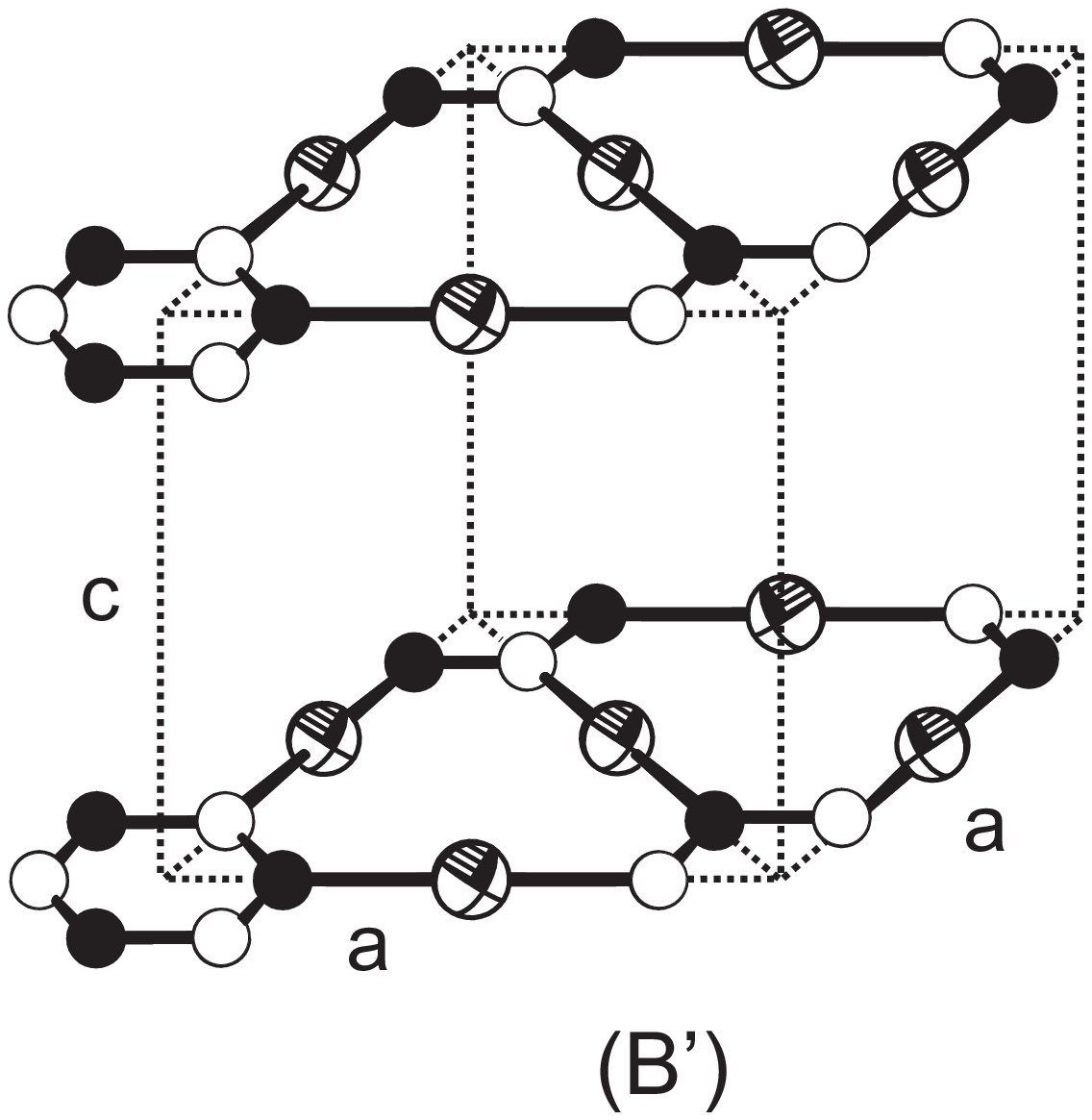}
\caption{Investigated crystal structures for Group 11 metal cyanides
($M$: spheres, C: black circles, N: white circles): A' - chain
structure $M$CN ($P6mm$), B' - sheet structure $M_3$C$_3$N$_3$ ($P\overline{6}2m$).}
\label{Fig:(A')(B')}
\end{figure}

One of the fundamental goals of
computational quantum chemistry is the prediction of new chemical
species and the determination of their
properties\cite{Gagliardi:06}. A recent example is gold cyanide,
AuCN, a well-known commodity chemical, whose known structure
consists of infinite -CN-Au-CN-Au- chains\cite{Zhdanov:01}. The
chains are packed together on a hexagonal grid in such a way, that
the Au atoms are in the same plane (see Fig. \ref{Fig:(A')(B')}:
A'). However, Hakala and Pyykk\"o\cite{Hakala:06} recently predicted
for AuCN an alternative crystal structure, having a closely similar
energy but only $\sim{ 70}$\% of the density (see Fig.
\ref{Fig:(A')(B')}: B'). The new structure contains triazine-type
six-rings of three carbon and three nitrogen atoms, C$_3$N$_3$,
which are coupled to each other by linearly coordinated gold(I)
atoms forming a two-dimensional sheet structure. The sheets attract
each other weakly due to the gold-gold aurophilic interaction. Such
a new structure has not yet been observed. Since single-sheet-type
2D atomic crystals, like graphene, are potentially a very important
class of materials but much less known than the 3D
counterparts\cite{Novoselov:05}, it is interesting to ask whether
such a structure could also be found for other similar systems, such
as the cyanides involving the other Group 11 (coinage) metals,
namely copper and silver. Their known crystalline structure is also
chain-type: -CN-$M$-CN-$M$- ($M$ denoting the
metal)\cite{Bowmaker,Hibble:02,Hibble:03b}.

By applying the valence isoelectronic principle, one can extend the
search for such a new sheet-type structures to further cases.
Replacing the formula $M$CN of the metal cyanides by the formula
$M$C$_2$, with an (ns)$^2$ configuration for the neutral metal atom
$M$, leads to the class of crystalline dicarbides. Metallic
dicarbides may have interesting electronic properties, such as the
superconductivity of YC$_2$ and its layered
compounds\cite{YC2-Henn}. In the present work we have studied
metallic elements in Groups 2 (alkaline earth metals $M$=Be, Mg, Ca,
Sr, Ba), and 12 (transition metals $M$=Zn, Cd and Hg). Stable
crystalline structures are known only for some of these systems
(MgC$_2$, CaC$_2$, SrC$_2$, BaC$_2$). No earlier data were found for
any Group 12 dicarbides.  The previously known crystal structures of
CaC$_2$, SrC$_2$, BaC$_2$ from Group 2 consist of infinite
-$M$-CC-$M$-CC- chains, but each cation is now equatorially
surrounded by four anions, each parallel to the $z$ axis (see Fig.
\ref{Fig:(A)(B)(C)}: B). Another known Group 2 dicarbide, MgC$_2$,
also has infinite chains, but the four equatorial dicarbide ligands
around the cation are now perpendicular to the $z$ axis (see Fig.
\ref{Fig:(A)(B)(C)}: C). These structures have been only very
recently clarified by X-ray and neutron
diffraction\cite{MgC2-Karen,SrC2-Vohn,BaC2-Vohn}.

In the present work we show that sheet-type structures such as the
poly(triaurotriazine), one recently reported\cite{Hakala:06} for
AuCN can be also found for the other Group 11 coinage metal cyanides
and for the Group 2 and 12 dicarbides. All the proposed sheet
structures seem to be entirely new. For Group 12 dicarbides and for
BeC$_2$ of Group 2 we also investigate whether they could form
stable structures in the tetragonal geometry starting the search
from the experimental geometry known for Group 2. We have performed
density of states (DOS) analysis for both isolated and packed
structures, and discuss the insulator vs.  metallic properties of
the studied systems. Vibrational analysis has been performed to
provide tools for the identification and to analyze the stability of
the systems. The geometries and frequencies are compared to the
known cyanide and dicarbide results. Finally, the thermodynamical
preferences of the predicted dicarbides are estimated by calculating
their formation energies with respect to the elemental metals and
graphite as starting materials.

\section{Method}

All calculations were performed within density functional theory
(DFT), using the Vienna {\it ab-initio} Simulation
Package\cite{VASP:01,VASP:02,VASP:03} (VASP). Plane-wave basis sets,
Ultrasoft (US) pseudopotentials\cite{US:01,US:02} and projector
augmented-wave (PAW) potentials\cite{PAW:01,PAW:02} with the
generalized gradient approximation\cite{PAW-GGA} (GGA) for the
Perdew-Burke-Ernzerhof\cite{PBE} (PBE) exchange-correlation
functional, were employed . When optimizing the lattice parameters
and the atomic positions, constant-volume calculations were
performed with the symmetry of the unit cell kept fixed. We
typically scanned the crystal volume for each system by changing the
interchain or intersheet distance in steps of 0.1~\AA~and letting
the system relax. In this work we will report the energy value and
the atomic geometry corresponding to the minimum point on the total
energy curve for each system.

For the energy cutoff we used a standard 400~eV value. A
$\Gamma$-centered 6$\times$6$\times$6 k-point grid was used
throughout the work for the lattice optimizations and vibrational
analysis. All the chain structures and the majority of the sheet
structures studied were found to be insulators, and this choice of
the k-point grid was found sufficient for the convergence in total
energy and geometry. For the dynamical matrix and vibrational
frequencies a shift of $0.025$~\AA~was applied to the atomic
positions. Some sheet structures exhibited the closing of the band
gap as the sheets were packed together, and two dicarbides were
manifestly metallic already as isolated sheets. The semimetallic
vs.~metallic character for these systems was inferred by studying
the DOS at the Fermi level. For the systems with a small band gap
the used k-point grid may lead to a slight overestimation of the
total energy ($\sim$0.005 eV) and the sheet-sheet distance
($\sim$0.2~\AA), as compared with the calculations using denser
k-point grids. We expect a similar behavior also for the two
metallic systems. In our tests, the intrachain and intrasheet
distances C-N, $M$-C and $M$-N were well converged within 0.2\% with
respect to increasing the k-point grid above that of
6$\times$6$\times$6. When the dispersion interactions are the
dominant mechanism, such as in the case of interchain distances in
cyanides, the present DFT-GGA results are not expected to be
reliable and should be considered with caution.

\begin{figure}[h]
\includegraphics[width=3.5cm,angle=-90,scale=1.00]{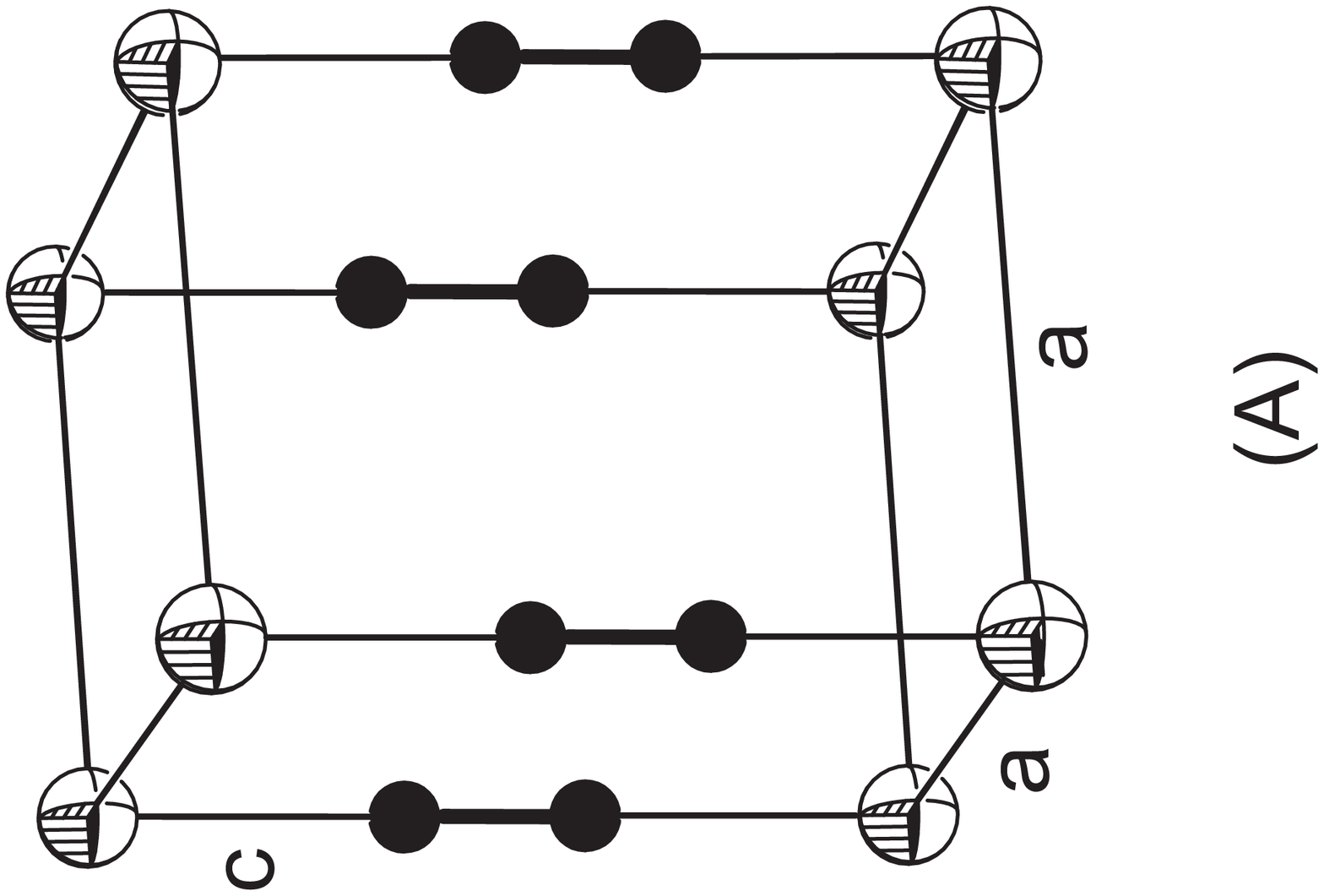}\hspace{4.5mm}
\includegraphics[width=3.5cm,angle=-90,scale=1.00]{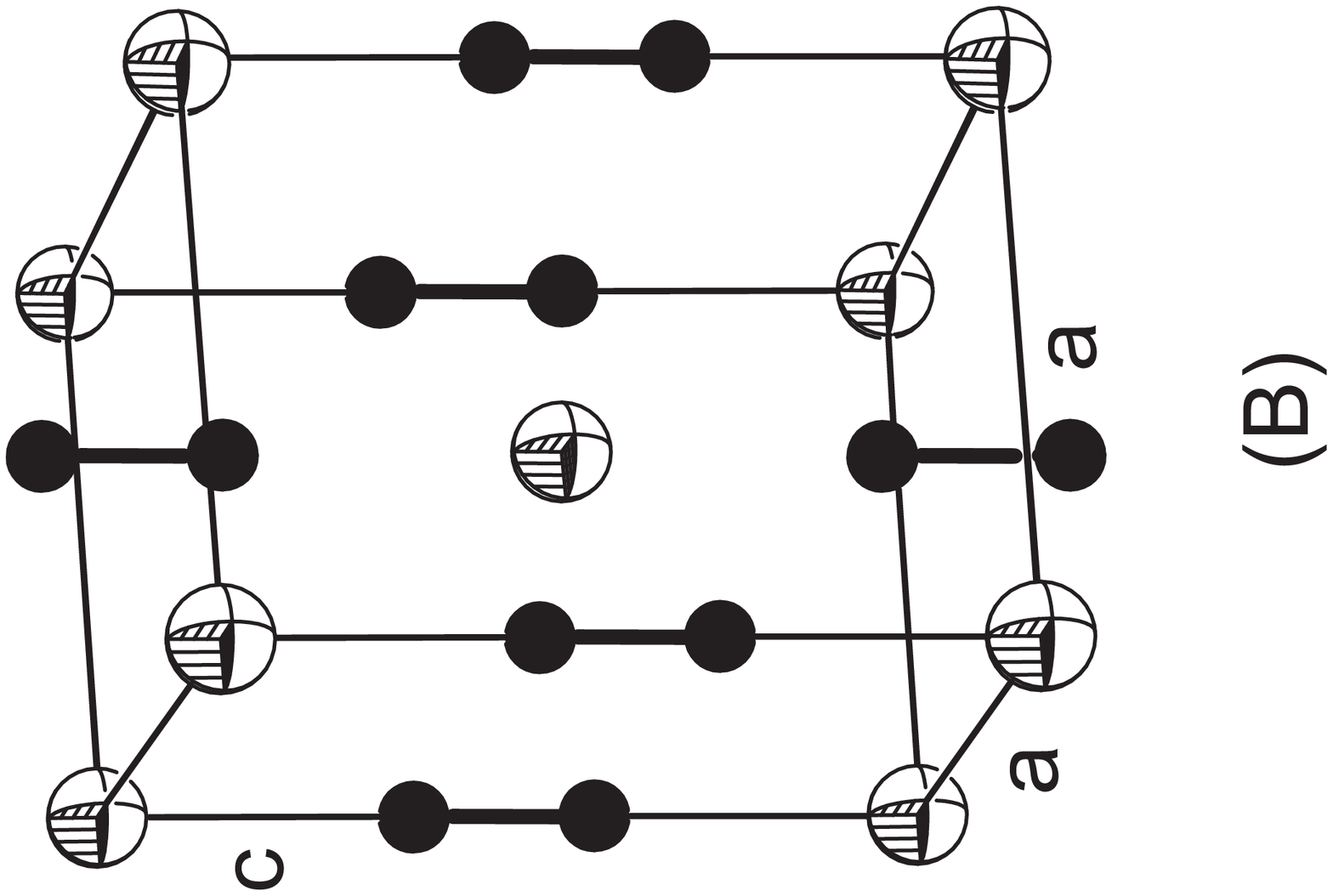}\hspace{3.5mm}
\includegraphics[width=3.5cm,angle=-90,scale=1.00]{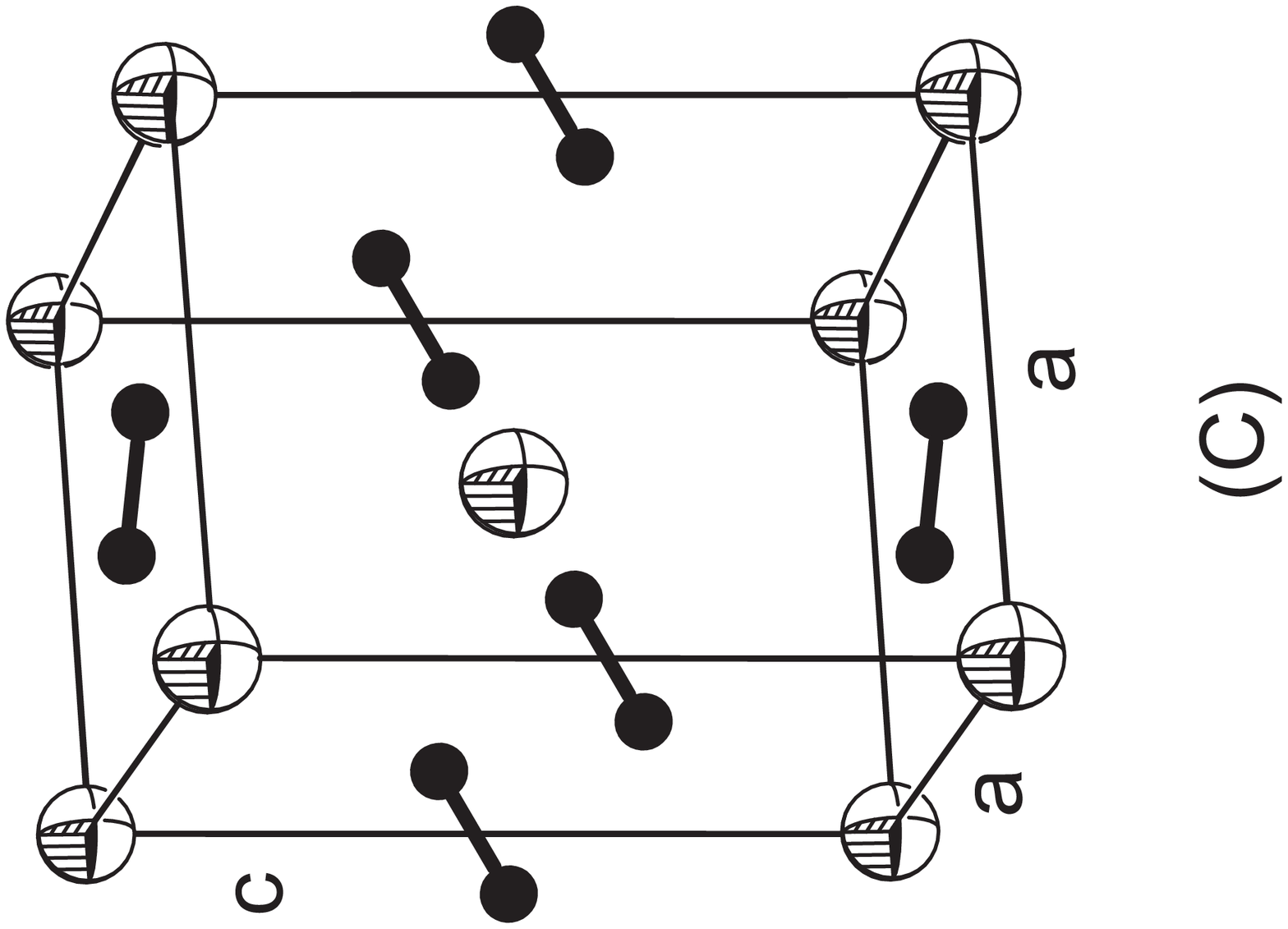}
\caption{Investigated crystal structures for chain-type metal dicarbides $M$C$_2$
($M$: spheres, C: black circles): A - tetragonal (P4/mmm), B -
tetragonal (I4/mmm), C - tetragonal
(I4$_2$/mnm).} \label{Fig:(A)(B)(C)}
\end{figure}

\begin{figure}[h]
\includegraphics[width=5.5cm,angle=-90,scale=0.70]{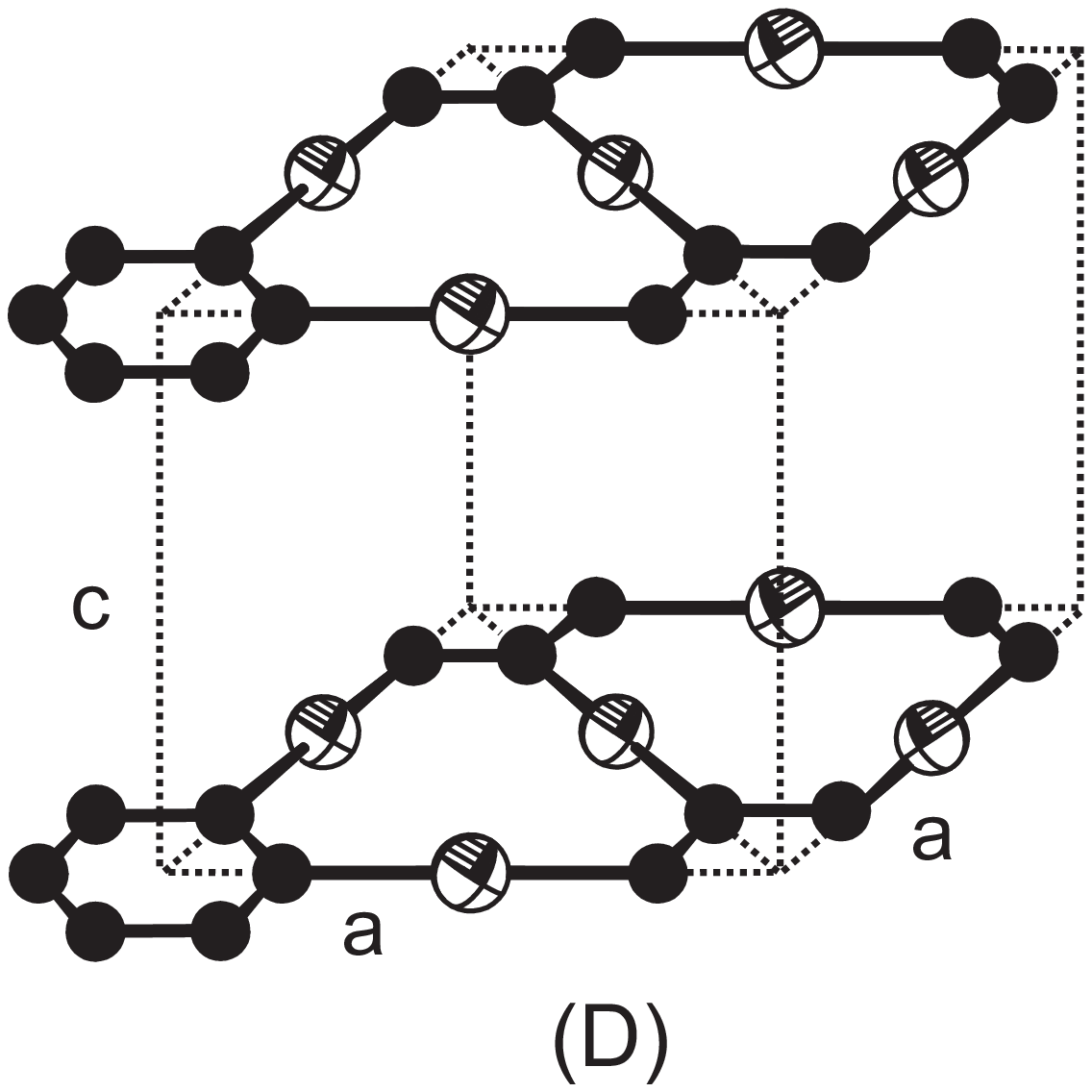}\hspace{0mm}
\includegraphics[width=5.5cm,angle=-90,scale=0.70]{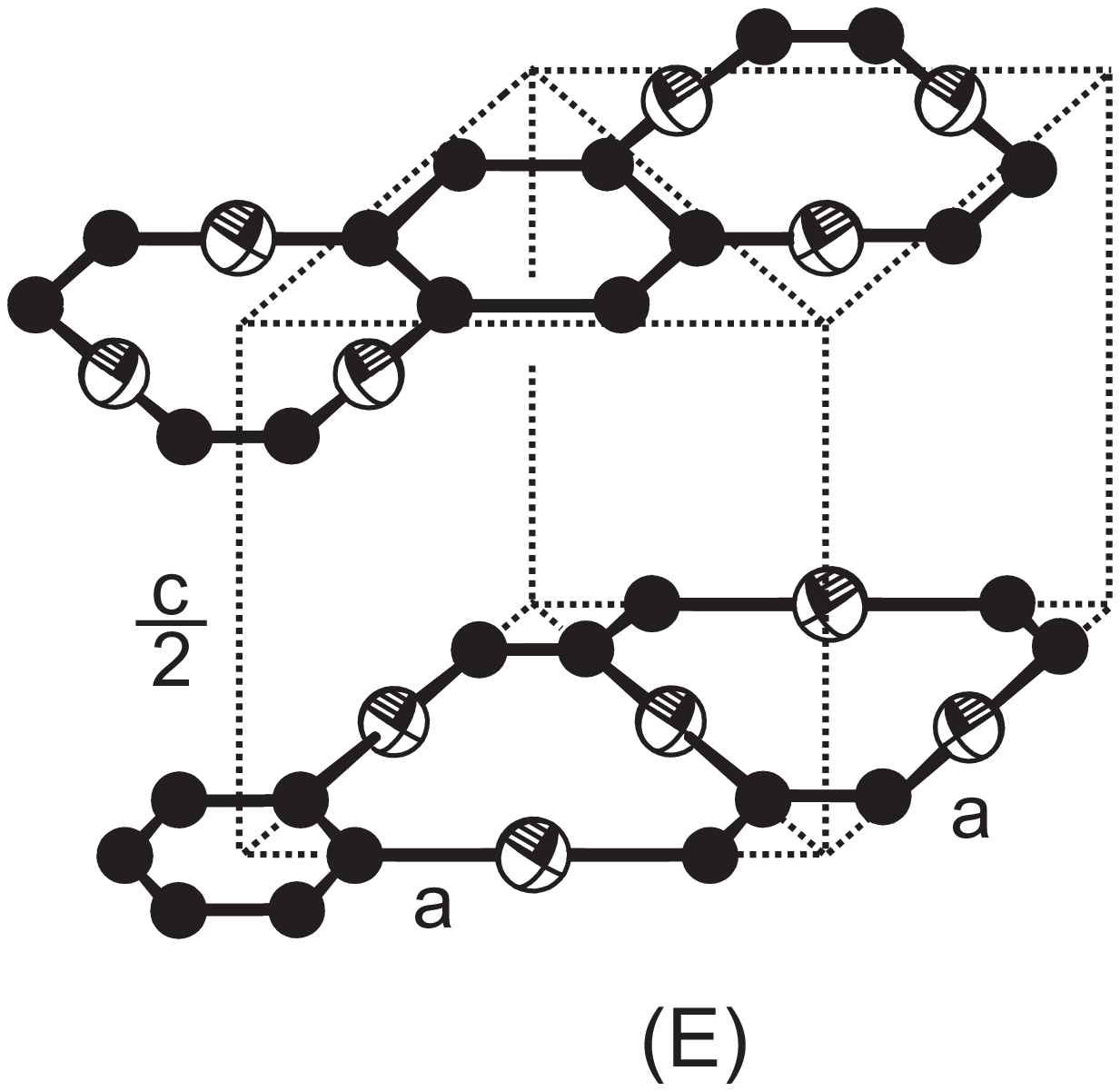}
\caption{Investigated crystal structures for sheet-type metal
dicarbides $M_3$C$_6$ ($M$: spheres, C: black circles): D - simple
stacking ($P{6\over m} m m$), E - shifted ($P\overline{6}2m$).}
\label{Fig:(D)(E)}
\end{figure}

\begin{figure}[h]
\includegraphics[width=5.5cm,angle=-90,scale=0.70]{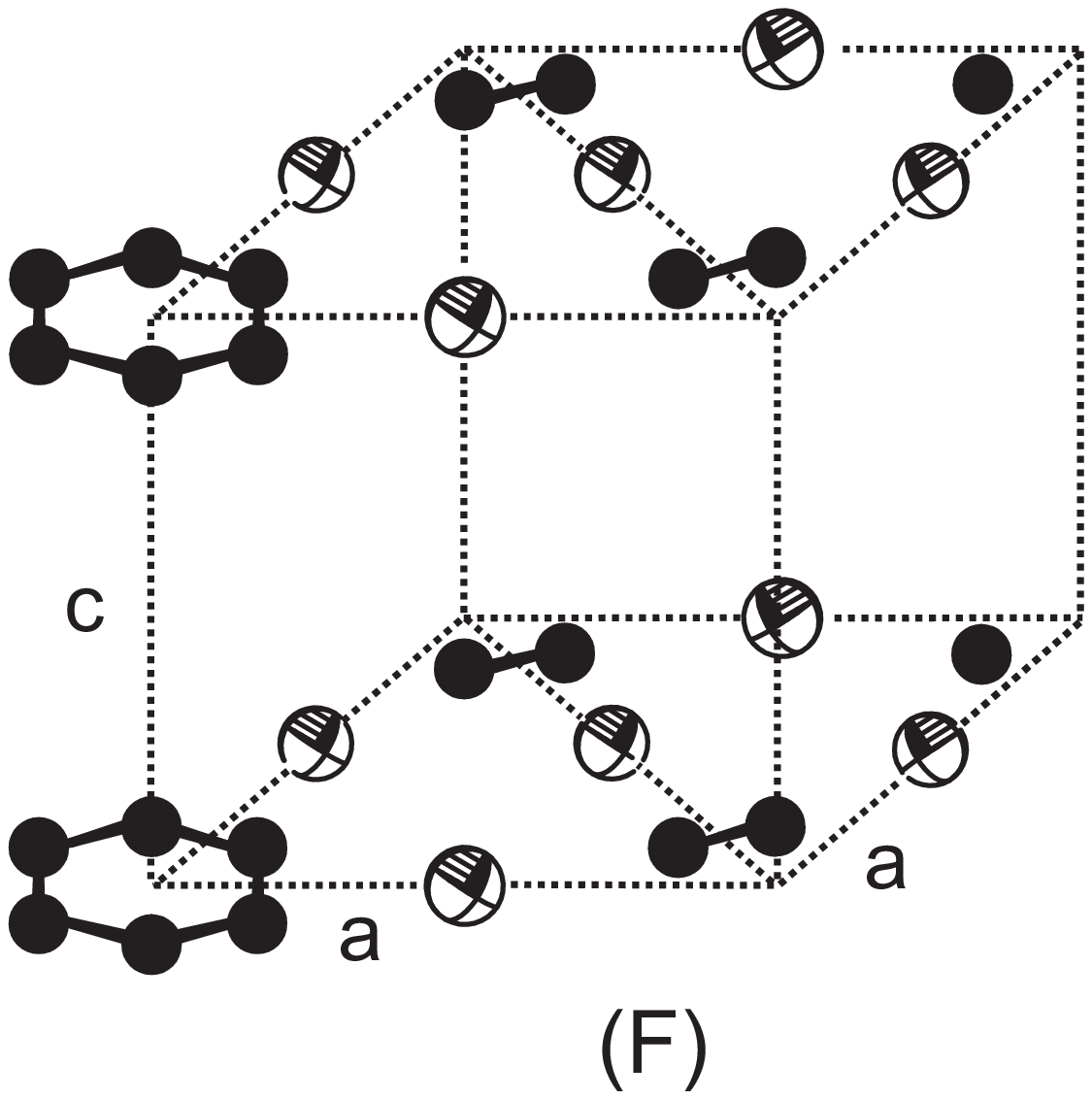}\hspace{0mm}
\includegraphics[width=5.5cm,angle=-90,scale=0.70]{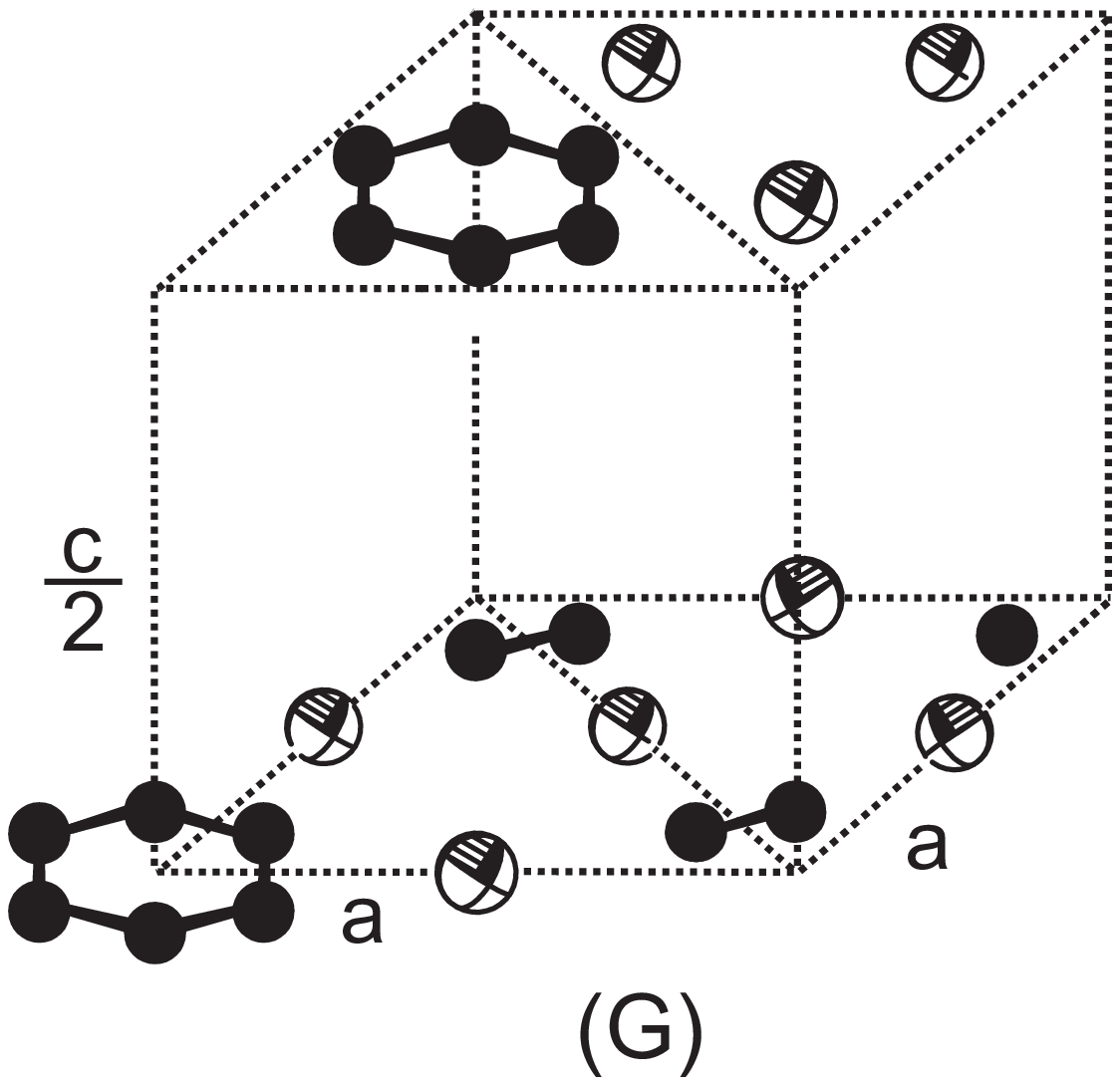}
\caption{Investigated crystal structures for sheet-type metal
dicarbides $M_3$C$_6$ ($M$: spheres, C: black circles): F - rotated
($P{6\over m} m m$), G - rotated plus shift ($P\overline{6}2m$).}
\label{Fig:(F)(G)}
\end{figure}

The studied geometries are illustrated in
Figs.~\ref{Fig:(A')(B')}-\ref{Fig:(F)(G)}. We refer to the
geometries as 'chains' (see Fig. \ref{Fig:(A')(B')}: A' and Fig.
\ref{Fig:(A)(B)(C)}: A, B, C), and to the rest as 'sheets'.
Furthermore, it turned out to be necessary to consider four
different packings of the sheets: 'Simple stacking', where each 2D
'sheet' is simply repeated in the perpendicular direction to form
the 3D crystal (see Fig. \ref{Fig:(A')(B')}: B' and Fig.
\ref{Fig:(D)(E)}: D); 'Shifted', where every second 2D sheet is
parallelly displaced, so that the benzene-like carbon ring lies on
top of the 'triangle' formed by the metal atoms (see Fig.
\ref{Fig:(D)(E)}: E); 'Rotated', as 'simple stacking', but the
carbon ring is rotated inside the sheet with respect to the
connecting metals by 30$^\circ$ (see Fig. \ref{Fig:(F)(G)}: F);
'Shifted+rotated', where in addition to the 'rotated' structure,
there is also the displacement parallel to the plane as described
above (see Fig. \ref{Fig:(F)(G)}: G).

\section{Results and Discussion}

\subsection{Geometries and total energies}

\subsubsection{Cyanides $M$CN}

The properties of the studied coinage metal cyanides CuCN, AgCN and
AuCN are found to be very similar concerning geometries and total
energies.  Like in the case of AuCN,\cite{Hakala:06} , our
simulations reproduce, for the two other compounds, the
experimentally known geometry within the chain and predict the
possible existence of a sheet-type structure. The chain-type
structure of AgCN and CuCN is known to differ from that of AuCN due
to an even weaker chain-chain interaction. In the case of AuCN the
aurophilic interaction keeps the interchain structure to a large
extent ordered (6-fold $P6mm$ symmetry about the chain axis),
whereas in the former cases there is a considerable degree of
disorder, leading to a lower 3-fold
symmetry.\cite{Bowmaker,Hibble:02,Hibble:03b} For the present
analysis all the chain-type structures were confined to the $P6mm$
symmetry (Fig.~\ref{Fig:(A')(B')}: A'). This is well sufficient for
the description of the strong covalent bonds C-N, $M$-C, and $M$-N
within the chains. The interchain search geometry for CuCN and AgCN
is not the one experimentally suggested but, as the attraction
between the chains is due to dispersion, DFT-GGA is not expected to
perform correctly for this distance in any case.

For the $M_3$C$_3$N$_3$ sheet structures, a total energy minimum was
found for the 'simple stacking' geometry (see
Fig.~\ref{Fig:(A')(B')}: B'), with symmetry $P\overline{6}2m$,
whereas no stable structures were found corresponding to the other
sheet geometries. The total energies for the chain and sheet
structures are compared in Table~\ref{Table:(Cyanides - Energy
difference)}.  The energy differences are about 0.2 eV for CuCN and
AgCN, but for AuCN there is practically no difference between the
two structures. Due to the weak sheet-sheet and chain-chain
interactions, the comparison of the isolated structures leads to the
same conclusion.

\begin{table}[h!]
\caption{Total energy difference {\it $\Delta$ E} (eV) between the
chain and sheet structures for Group 11 metal cyanides (per $M$CN
formula unit). Both 3D and isolated cases are reported.  Negative
value indicates that the chain structure is more stable.}
\begin{ruledtabular}
\begin{tabular}{lccc}
    Case                  &  CuCN          &   AuCN          &       AuCN   \\
\hline
    3D Sheet vs.~3D Chain &  -0.22         &  -0.18          &       0.00   \\
    2D Sheet vs.~1D Chain &  -0.24         &  -0.21          &       0.03
\end{tabular}
\end{ruledtabular}
\label{Table:(Cyanides - Energy difference)}
\end{table}

Table~\ref{Table:(Cyanides - bond lengths)} gives the bond lengths
and the lattice constants, along with the experimental values for
the known $M$-N, $M$-C and C-N distances. We compare our results
with the experimental values from Hibble and
co-workers\cite{Hibble:02,Hibble:03,Hibble:03b}, who performed a
systematic total neutron diffraction study for these structures.
Their results differ somewhat from the conventional Bragg scattering
studies, which, according to them, do not yield completely
satisfactory results. For the chain structures, the experimental
intrachain covalent bond lengths are very well described with
distances accurate to ~2\% or better. In particular, the C-N
distance is well reproduced. The difference between the $M$-C and
$M$-N bond lengths in sheets compared to the same difference in
chains, is larger. For sheets, the C-N bond lengths are practically
the same for all the cyanides, but larger by about 20 pm compared to
the chain structure, reflecting the aromatic six-ring bonding, as
contrasted to the -C$\equiv$N- triple bonds.

Because of the well known difficulties in describing dispersion
effects by DFT, the interchain and intersheet distances should be
considered much more uncertain, which can be seen in the large
theory-experiment discrepancy of the lattice parameter $a$ for AuCN,
which for this system corresponds to the chain-chain distance. The
comparison of the interchain distances with the experimental values
is further complicated, because the chains are randomly displaced
along the chain axis, and for AgCN and CuCN there are no chemical
bonds between the chains\cite{Hibble:02,Hibble:03,Hibble:03b}. As a
matter of fact, the latter is reflected in the very weak binding
energy between the isolated and packed CuCN and AgCN chains in our
calculations. The binding energy for isolated sheets required to
form a 3D structure is found to be typically 0.05~eV.

It is interesting to note that room-temperature powder-X-ray pattern
for CuCN shows unexplained weak Bragg peaks at distances
$d_1=260.5$~pm~and $d_2=235.3$~pm~(Ref.\onlinecite{Hibble:03b}).
These have been attributed to an unknown highly crystalline form
$\beta$-CuCN with density 2.97 g~cm$^{-3}$ (cf.~3.03 g~cm$^{-3}$ for
the known $\alpha$-CuCN). Whether such peaks could be correlated
with our predicted sheet-type geometry or its modifications remains
an open question. The crystal density of the sheet structure is 1.8
g~cm$^{-3}$, i.e., $\sim 40$\% smaller than that assigned to
$\beta$-CuCN. This computed density is, however, strongly influenced
by the sheet-sheet distance $c$, which is possibly underestimated
due to the lack of dispersion interaction. The present value of $c$
is $\sim 28$\% and $\sim 41$\% larger than the distances $d_1$ and
$d_2$, respectively.

\begin{table*}[h!]

\caption{Bond lengths, lattice parameters $a$ and $c$ (pm) and volumes $V$ (\AA$^3$)
for Group 11 metal cyanides. The volume is given with respect to one
$M$CN unit. The values $a$ and $V$ given in italics are not directly
comparable to the experiment due to the different crystal symmetry in
the calculation. The capital letter after the compound refers to the
crystal structures in Fig.~\ref{Fig:(A')(B')}.}
\begin{ruledtabular}
\begin{tabular}{cccccccc}
  Chains  & Source                                      &  $M$-C    &  $M$-N     &    C-N     &   $a$           &   $c$         & $V$          \\
\hline
CuCN (A') &Calc.                                        &  182.0    &  181.9     &   117.4    &   {\it 407.7}   &   481.2       & {\it 69.27}  \\
          &Exp.\footnote{Ref.~\onlinecite{Hibble:03b}}  &  184.6    &  184.6     &   117.0    &   591.2(3)      &   486.107(3)  & 49.0467      \\
AgCN (A') &Calc.                                        &  201.9    &  204.0     &   117.0    &   {\it 410.8}   &   523.9       & {\it 76.56}  \\
          &Exp.\footnote{Ref.~\onlinecite{Hibble:02}}   &  206      &  206       &   116      &   590.32        &   528.29      & 53.1443      \\
AuCN (A') &Calc.                                        &  194.3    &  198.3     &   116.8    &   379.4         &   511.3       & 63.74        \\
          &Exp.\footnote{Ref.~\onlinecite{Hibble:03}}   &  197.03(5)& 197.03(5)  &  114.99(2) &   340.5(4)      &   509.2(2)    & 51.1273      \\
\hline
    Sheet  & Source                                     &  $M$-C    &  $M$-N     &    C-N     &   $a$           &   $c$         & $V$          \\
\hline
    Cu$_3$C$_3$N$_3$ (B') &Calc.                        &  187.6    &  190.0     &   136.6    &   650.7         &   332.4       & 81.24        \\
    Ag$_3$C$_3$N$_3$ (B') &Calc.                        &  207.4    &  213.8     &   136.1    &   693.3         &   341.5       & 94.78        \\
    Au$_3$C$_3$N$_3$ (B') &Calc.                        &  199.5    &  210.4     &   136.3    &   682.5         &   352.4       & 94.78
\label{Table:(Cyanides - bond lengths)}
\end{tabular}
\end{ruledtabular}
\end{table*}

\subsubsection{Dicarbides $M$C$_2$}
\label{Sec:Dicarbides}

The geometries of the metal dicarbides in the known experimental
chain structure ($M$ = Mg-Ba in Group 2) are well reproduced by our
calculations. In addition, also for the dicarbides we predict the
possible existence of sheet structures for both Groups 2 and 12, and
for Group 12 we find possible chain structures. Experimentally, the
Group 2 dicarbides crystallize in two different tetragonal
symmetries, {\it viz} I4/mmm for $M$C$_2$ ($M$ = Ca-Ba) and
I4$_2$/mnm for MgC$_2$. Table \ref{Table:(Carbides - group 2 - bond
lengths chains)} contains the calculated bond lengths and lattice
parameters for the Group 2 chain dicarbides in comparison with the
available experimental data.  Table~\ref{Table:(Carbides - group 12
- bond lengths chains)} contains the predictions for chain
structures for the Group 12.

As most of the chain- and sheet-type dicarbide structures were found
to have no or low imaginary frequencies, the structures can be
considered stable. The frequencies are presented in
Sec.~\ref{vibrational_frequencies}. Notable exceptions are BeC$_2$
and Be$_3$C$_6$, which exhibited imaginary frequencies of the order
of 300$i$~cm$^{-1}$, suggesting that the found minimum energy
geometry is only a transition state between phases of other
symmetry. We did not try to follow the mode to a lower-symmetry
minimum. Some imaginary frequencies of the order of
100$i$~cm$^{-1}$, were found for chain-type MgC$_2$, whose structure
is, however, in a very good agreement with the experiment. Smaller
imaginary frequencies (about 50$i$~cm$^{-1}$ or smaller) were found
for ZnC$_2$, CdC$_2$, Ca$_3$C$_6$, Sr$_3$C$_6$ and Ba$_3$C$_6$. We
believe the origin of these imaginary frequencies is related either
to the numerical inaccuracies or the anharmonicity of the total
energy surface. In fact, small imaginary frequencies have been
interpreted in terms of soft mode phonons.\cite{Sternik:01}

For the known Group 2 dicarbides in the chain structure, the
calculated covalent bond lengths are in good agreement with
experiment. Our calculations systematically give slightly larger
values of the C-C bond length (around 125~pm) compared to the
experimental results (around 120~pm), the C-C bond length in
acetylene molecule\cite{Herzberg} (120.4~pm) or the bond length
calculated from the carbon triple-bond covalent radius
(120~pm)\cite{Pyykko:05}. A possible reason for the difference could
be the vibrational motion of the dicarbide group in the
experiments\cite{MgC2-Karen}. In realistic physical conditions, when
thermal energy is applied, C$_2^{2-}$ group can both rotate and
vibrate, thus leading to an underestimation of the C-C bond length
by experimental measurement. An interesting point to notice is that
for ZnC$_2$, CdC$_2$ and HgC$_2$, we predict the possible existence
of a chain structure in the I4/mmm geometry similar to the $M$C$_2$
($M$ = Ca-Ba) structures. The I4$_2$/mnm geometry for ZnC$_2$,
CdC$_2$ and HgC$_2$ was also found to be stable, but with slightly
higher energies.

The difference in the packing geometry of the chain-type dicarbides
compared to the packing geometry of the coinage metal cyanides can
be understood from the different relative contributions to the
interaction forces. Between the monovalent (Au$^+$)(CN$^-$) chains
the dispersion forces predominate and one Au$^+$ is surrounded by
six other nominally Au$^+$ ions. For the divalent
($M^{2+}$)(C$_2^{2-}$) chains, on the other hand, the Coulomb forces
predominate. Hence in this case the ($M^{2+}$)(C$_2^{2-}$) chains
are displaced with respect to each other, so that one $M^{2+}$ ion
is surrounded by four C$_2^{2-}$ ions (see Fig. \ref{Fig:(A)(B)(C)}:
B, C). In support of this, infinite ($M^{2+}$)(C$_2^{2-}$) chains
packed into a tetragonal P4mm unit cell (see Fig.
\ref{Fig:(A)(B)(C)}: A) were found purely repulsive.

\begin{table*}[h!]
\caption{Experimental and calculated lattice parameters $a$ and $c$
(pm), cell volumes $V$ (\AA$^3$), symmetries and bond lengths (pm)
for Group 2 metal dicarbides in the chain structure. Volume is given
with respect to one $M$C$_2$ unit. The capital letter after the
compound refers to the crystal structures in Fig.~2.}
\begin{ruledtabular}
\begin{tabular}{lcccccccccc}
Case                                       & BeC$_2$ (B)   & \multicolumn{2}{c}{MgC$_2$ (C)}  & \multicolumn{2}{c}{CaC$_2$ (B)} & \multicolumn{2}{c}{SrC$_2$ (B)} & \multicolumn{2}{c}{BaC$_2$ (B)}  \\
\cline{2-2}\cline{3-4}\cline{5-6}\cline{7-8} \cline{9-10}                              & Calc.         &  Calc.       &       Exp.\footnote{Ref. \cite{MgC2-Karen}}            &  Calc.          & Exp.\footnote{Ref. \cite{CaC2-Knapp}}        &  Calc.        & Exp.\footnote{Ref. \cite{SrC2-Vohn}}        &  Calc.        &  Exp.\footnote{Ref. \cite{BaC2-Vohn}}         \\
\cline{1-10}
        Lattice const.: $a$                & 390.6         &  396.9       &  393.42(7)$^a$    &  391.3        & 388.582(4)$^a$  &  414.4        & 411.43(2)$^a$   &  440.6       &  439.43(6)$^a$    \\
\hspace{2.075cm}        $c$                & 463.2         &  492.5       &  502.1(1)         &  640.0        & 640.05(1)       &  681.8        & 676.60(4)       &  724.1       &  712.5(2)         \\
     Volume $V$                            & 35.01         &  38.8        &  38.85            &  49.0         & 48.20           &  58.60        & 57.26           &  70.3        &  68.88            \\
Symmetry                                   & I4/mmm        &\multicolumn{2}{c}{I4$_2$/mnm}    & \multicolumn{2}{c}{I4/mmm}      & \multicolumn{2}{c}{I4/mmm}      & \multicolumn{2}{c}{I4/mmm}       \\
\hline
Bond lengths                               &               &              &                   &               &                 &               &                 &               &                  \\
\hline
\hspace{4.5mm}{Chain-chain dist.}          & 276.1         &  280.6       &  278.2            &  276.7        & 274.7           &  293.0        & 290.8           &  311.6        &  310.7           \\
\hspace{6mm}{\it M}-C                      & 169.1         &  217.8       &  217.4            &  256.8        & 255.5(4)        &  277.7        & 278.5(8)        &  298.8        &  297.0(7)        \\
\hspace{6.3mm}C-C                          & 124.9         &  125.8       &  121.5            &  126.4        & 129.7(8)        &  126.4        & 120(2)          &  126.6        &  118.6(13)       \\

\end{tabular}
\end{ruledtabular}
\label{Table:(Carbides - group 2 - bond lengths chains)}
\end{table*}

\begin{table}[h!]
\caption{Calculated lattice parameters $a$ and $c$ (pm), cell
volumes $V$ (\AA$^3$), symmetries and bond lengths (pm) for Group 12
metal dicarbides in the chain structure. Volume is given with
respect to one $M$C$_2$ unit. The capital letter after the compound
refers to the crystal structures in Fig.~2.}
\begin{ruledtabular}
\begin{tabular}{llccc}
Case                                      &      & ZnC$_2$ (B) & {CdC$_2$} (B)    & {HgC$_2$} (B)     \\
\hline Lattice constants:                 & $a$  & 425.1     & 438.1        & 510.5         \\
                                          & $c$  & 505.2     & 546.9        & 529.0         \\
     Volume $V$                           &      & 45.7      & 52.5         & 68.9          \\
Symmetry                                  &      & I4/mmm    & I4/mmm       & I4/mmm        \\
\hline
Bond lengths                              &      &           &              &               \\
\hline
\hspace{4.5mm}{Chain-chain dist.}         &      & 300.6     & 309.8        & 361.0         \\
\hspace{6mm}{\it M}-C                     &      & 190.5     & 211.2        & 202.9         \\
\hspace{6.3mm}C-C                         &      & 124.2     & 124.5        & 123.2         \\
\end{tabular}
\end{ruledtabular}
\label{Table:(Carbides - group 12 - bond lengths chains)}
\end{table}

\begin{table*}[h]
\caption{Calculated lattice parameters $a$ and $c$ (pm), volumes $V$
(\AA$^3$), symmetries and bond lengths (pm) for Group 2 and 12 metal
dicarbides in the sheet structure. Volume is given with respect to
one $M$C$_2$ unit. The capital letter after the compound refers to
the crystal structures in Figs.~\ref{Fig:(D)(E)}-\ref{Fig:(F)(G)}.}
\begin{ruledtabular}
\begin{tabular}{lcccccccc}
Case                                        &Be$_3$C$_6$ (E)& Mg$_3$C$_6$ (G)  &Ca$_3$C$_6$  (G)   & Sr$_3$C$_6$  (G)  & Ba$_3$C$_6$  (G) & Zn$_3$C$_6$  (E)  &Cd$_3$C$_6$  (E) & Hg$_3$C$_6$  (E)  \\
\hline
        Lattice const.: $a$                 & 390.6         &  665.0     &  736.2     &  782.7        & 826.2       &  674.4        & 715.2   &  708.2       \\
\hspace{2.075cm}        $c$                 & 463.2         &  526.8     &  543.3     &  570.0        & 592.1       &  711.1        & 695.1   &  794.1       \\
     Volume $V$                             & 56.9          &  67.2      &  84.9      &  100.7        & 116.9       &  93.4         & 102.6   &  115.1       \\
Symmetry                                    & $P\overline{6}2m$ &   $P{6\over m} m m$      &  $P{6\over m} m m$     &   $P{6\over m} m m$         &  $P{6\over m} m m$        &  $P\overline{6}2m$         &  $P\overline{6}2m$    &  $P\overline{6}2m$        \\
\hline
Bond lengths                                &               &            &            &               &            &               &         &              \\
\hline
\hspace{6mm}{\it M}-C                       & 168.9         &  214.2     &  246.2     &  273.2        & 280.4     &  194.8        & 214.9     &  212.7       \\
\hspace{6.3mm}C-C                           & 142.9         &  142.9     &  143.2     &  140.0        & 141.2     &  142.6        & 141.9     &  140.7       \\

\end{tabular}
\end{ruledtabular}
\label{Table:(Carbides - group 2+12 - bond lengths sheets)}
\end{table*}

In the search of stable 2D infinite metal dicarbide sheets
$M_3$C$_6$, the similar hexagonal symmetry D as in the
$M_3$C$_3$N$_3$ systems B' was used as a starting point (initially
with a large sheet separation). Here, the benzene-like carbon rings
are coupled together from the corners by the metal atoms. The
calculations in this geometry yielded minimum energy structures for
$M$ = Be, Zn-Hg. For the remaining metals, the isolated 2D sheet
spontaneously relaxed to a slightly different geometry within the
same hexagonal crystal symmetry. The carbon rings rotated 30$^\circ$
with respect to the initial structure to yield the 2D sheet geometry
shown in F and G. In this structure the metal atoms are located in
the midpoint of the line connecting the C-C edges of every two
benzene-like rings. Considering the 3D stacking of these structures,
a 'simple stacking' (both 'rotated' and 'nonrotated' cases) was
found purely repulsive. Instead, a new 3D packing had to be
introduced, where every second sheet was translated by the vector
$\vec{v}= ({1\over 2}a, {1\over 2}a, 0)$. For both the structures (E
and H), stable geometries were found.

\begin{figure}[h]
\includegraphics[height=11cm,angle=-90,scale=0.78]{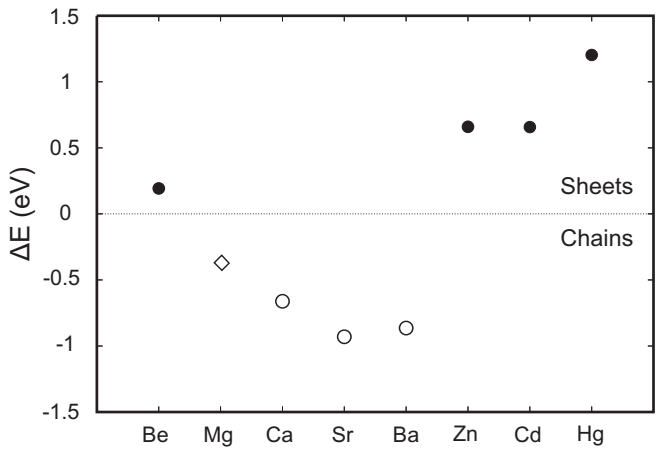}
\caption{Total energy differences (eV) between the most stable chain
and sheet structures for Group 2 and 12 metal dicarbides: Chain B -
Sheet F ($\bullet$); Chain B - Sheet H ($\circ$); Chain C - Sheet F
($\diamond$).} \label{Fig:(Plot-All)}
\end{figure}

Fig.~\ref{Fig:(Plot-All)} summarizes the behavior of total energies
for the most stable sheets and chains (the numerical values are
tabulated later on in Sec.~\ref{sec:formation}). The energy
difference {\it $\Delta E$} is given for each dicarbide per $M$C$_2$
unit; a negative value indicates that the chain structure is
preferred. For the experimentally known dicarbides, $M$ = Mg-Ba, the
chains are energetically more stable. The energy difference roughly
increases following $Z$, being in all the cases less than 1 eV.
Interestingly, for the dicarbides not experimentally known, $M$ =
Be, Zn, Cd and Hg, the sheet structure is predicted to be the most
stable phase, and the energy difference likewise increases roughly
following $Z$.

\subsection{Vibrational frequencies}
\label{vibrational_frequencies}

\begin{figure}[h]
\includegraphics[height=8.6cm,angle=-90,scale=1.00]{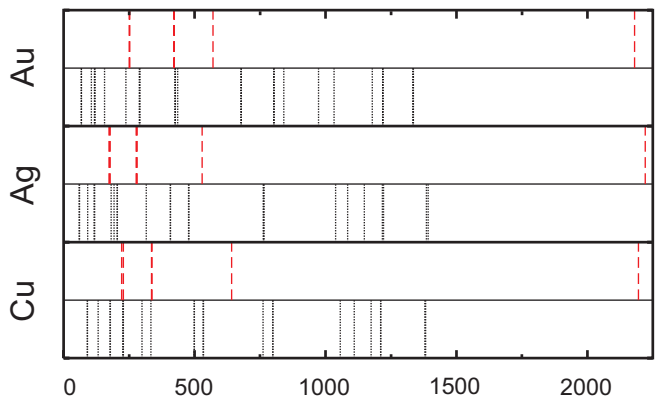}
\caption{Calculated vibrational frequencies (cm$^{-1}$) for Group 11
metal cyanides. Dotted lines (black) refer to the sheet-type
structures, dashed lines (red) refer to the chain-type structures.
For comparison with experiment, see Table \ref{Table:(Cyanides -
Vibrations)}.} \label{Fig:vibr_cyanides}
\end{figure}

\begin{figure}[h]
\includegraphics[height=8.6cm,angle=-90,scale=1.00]{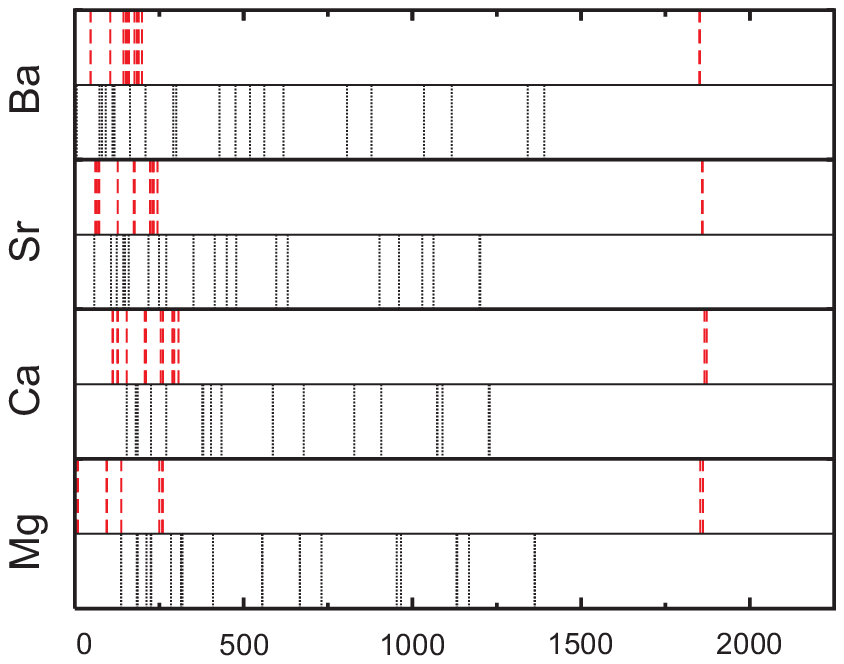}
\caption{Calculated vibrational frequencies (cm$^{-1}$) for Group 2
metal dicarbides. Dotted lines (black) refer to the sheet-type
structures, dashed lines (red) refer to the chain-type structures.
For comparison with experiment, see Table \ref{Table:(Carbides -
Vibrations)}.} \label{Fig:vibr_dicarb2}
\end{figure}

\begin{figure}[h]
\includegraphics[height=8.6cm,angle=-90,scale=1.00]{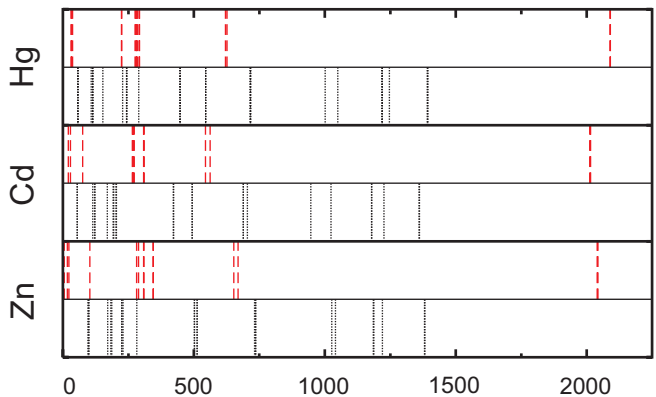}
\caption{Calculated vibrational frequencies (cm$^{-1}$) for Group 12
dicarbides. Dotted lines (black) refer to the sheet-type structures,
dashed lines (red) refer to the chain-type structures.}
\label{Fig:vibr_dicarb12}
\end{figure}

For the cyanide structures, vibrational analysis was performed for
isolated chains and isolated sheets, whereas for dicarbides the
analysis was done for packed chains and isolated sheets. The rest of
the packed structures were not studied because of the difficulties
due to the weak chain-chain (for cyanides) and sheet-sheet (for both
cyanides and dicarbides) interactions. The frequencies are given in
Figs.~\ref{Fig:vibr_cyanides}-\ref{Fig:vibr_dicarb12}, and
Tables~\ref{Table:(Cyanides - Vibrations)} and~\ref{Table:(Carbides
- Vibrations)},\ref{Table:(Carbides - Vibrations)2}  with comparison
to available
experiments.\cite{Bowmaker,MgC2-Karen,CaC2-Knapp,SrC2-Vohn,BaC2-Vohn}.

Following the assignment by Bowmaker~{\it et al.}\cite{Bowmaker} for
the modes in infinite cyanide chains, it can be seen that the
simulation reproduces very well both the vibrational and bending
normal modes for the known structures.  Here the two highest modes
are the $\nu$(CN) and $\nu$($M$C/N) stretching modes, and the two
lowest ones the $\delta$($M$CN) and $\delta$(N$M$C) bending modes.
For the known dicarbides in the chain structure, the highest
frequencies correspond to the stretching mode and are very close to
the experimental ones.

The sheet structures for both cyanides and dicarbides exhibit
systematically smaller maximum frequencies compared to those of the
chain structures. The higher eigenfrequencies for sheets correspond
to in-plane vibrations. Similarly as the bending modes in chains,
the majority of the lower frequencies in sheets corresponds to
bending vibrations perpendicular to the plane. For AuCN, the first
and second frequency (Table \ref{Table:(Cyanides - Vibrations)}) are
close to the in-plane ring deformations of an isolated molecule
(1395 and 1208cm$^{-1}$)~(Ref.~\onlinecite{Hakala:06}).

We propose that the possible experimental identification of the
sheet-type structures for cyanides could be performed within the
range of roughly (650-1500) cm$^{-1}$. The majority of the in-plane
vibrations of the sheets falls into this wavelength region, where no
frequencies corresponding to chains interfere.  Similarly, the
identification of sheet-type dicarbides could be done in the range
(700-1800) cm$^{-1}$, undisturbed by the frequencies of the known
chain structures. The frequencies in these ranges correspond mainly
to vibrations parallel to the sheets (i.e., parallel to the covalent
bonds), and are thus not significantly affected by the DFT
deficiencies for dispersion.

\begin{table*}[h!]
\caption{Experimental and calculated vibrational frequencies
(cm$^{-1}$) for Group 11 metal cyanides. For sheets, the four
highest frequencies, which correspond to in-plane stretching modes,
are reported. The close-lying frequencies have been grouped
together.}
\begin{ruledtabular}
\begin{tabular}{lcccccc}

Case                              & \multicolumn{2}{c}{CuCN}& \multicolumn{2}{c}{AgCN} & \multicolumn{2}{c}{AuCN}\\
\cline{2-3}\cline{4-5}\cline{6-7}
Chains                            & Calc.      &  Exp.\footnote{Ref. \cite{Bowmaker}}      & Calc.      &  Exp.$^a$       & Calc.      &  Exp.$^a$      \\
\hline
Stretch (C-N)                     & 2195       & 2170       & 2221       & 2164        & 2180        & 2236      \\
Stretch ($M$-CN)                  & 642        & 591        & 529        & 480         & 571         & 598       \\
Bend 1                            & 337,336    & 326        & 281,277    & 272         & 422,421     & 358       \\
Bend 2                            & 228,221    & 168        & 178,173    & 112         & 253,251     & 224       \\
\hline
Sheets                            & Calc.      &  Exp.      & Calc.      &  Exp.       & Calc.      &  Exp.      \\
\hline
                                  & 1382,1379  &  -         & 1392,1386  &   -         & 1336,1334               &   -         \\
                                  & 1212,1210  &  -         & 1221,1216  &   -         & 1219,1218               &   -         \\
                                  & 1174,1109,1057  &  -    & 1174,1087,1039       & -  & 1178,1032,973          &    -         \\
                                  & 799,798    &   -        & 763,762    &   -          & 804,802                &   -          \\

\end{tabular}
\end{ruledtabular} \label{Table:(Cyanides -
Vibrations)}
\end{table*}

\begin{table*}[h!]
\caption{Experimental and calculated vibrational frequencies
(cm$^{-1}$) for Group 2 and 12 chain-type dicarbides. The highest
frequencies, which correspond to stretching modes along the chains,
are reported. The close-lying frequencies have been grouped
together.}
\begin{ruledtabular}
\begin{tabular}{lcccccccc}
Chains                            & MgC$_2$         & CaC$_2$       & SrC$_2$        & BaC$_2$    &  ZnC$_2$    &  CdC$_2$  &  HgC$_2$   \\
\hline
Experimental value\footnote{Ref.~\onlinecite{Reckeweg01}}             & -              & 1860          & 1850           & 1832       & -           & -         & -          \\

Calculated value                  & 1861,1854       & 1872,1866     & 1861,1860      & 1853,1851  & 2043,2040   & 2015,2012  & 2090,2090 \\
                                  & 261,260         & 308,294       & 245,234        & 200,190    & 669,652     & 563,545    & 627,621   \\
\end{tabular}
\end{ruledtabular}
\label{Table:(Carbides - Vibrations)}
\end{table*}

\begin{table*}[h!]
\caption{Calculated vibrational frequencies (cm$^{-1}$) for Group 2
and 12 sheet-type dicarbides. Only the highest frequencies values,
which correspond to in-plane stretching modes, are reported. The
close-lying frequencies have been grouped together.}
\begin{ruledtabular}
\begin{tabular}{cccccccc}
       Mg$_3$C$_6$        & Ca$_3$C$_6$    & Sr$_3$C$_6$    & Ba$_3$C$_6$    & Zn$_3$C$_6$  & Cd$_3$C$_6$     & Hg$_3$C$_6$        \\
\hline
       1364,1362          & 1230,1227      & 1202,1199      & 1391,1342      & 1382,1381    & 1361,1360       & 1394,1391         \\
       1168               & 1090,1074      & 1063,1030      & 1117,1035      & 1220         & 1226            & 1246              \\
       1134,1131          & 1073           & 963,901        & 879,807        & 1187,1185    & 1179,1178       & 1219,1218         \\
       966,954            & 909,828        & 631            & 619            & 1040,1027    & 1024,947        & 1050,1001

\end{tabular}
\end{ruledtabular}
\label{Table:(Carbides - Vibrations)2}
\end{table*}

\subsection{Band gap and metallicity}
\label{sec:DOS}

The band gap values for the studied systems are reported in
Table~\ref{Table:(DOS - HOMO/LUMO gap)}. Since a full band structure
analysis was not performed, the values given should be considered as
indicative only. In the case of both cyanide and dicarbide chain
structures, all the known and predicted 1D (isolated) and 3D
structures are found to be insulators. For the packed 3D dicarbide
chain structures, the band gap is largest for HgC$_2$ (3.7 eV) and
smallest for BaC$_2$ (1.6 eV).  The packed chain structures exhibit
systematically smaller band gaps than the isolated chains due to the
increased overlap of the valence orbitals. The only exception to
this picture is BaC$_2$, which seems to show an opposite trend. All
the newly predicted chain dicarbides ($M$ = Zn-Hg) have larger gaps
than the already known chain dicarbides.

For both cyanide and dicarbide 2D (isolated) and 3D sheet structures
the band gaps are clearly smaller than for the chain structures.
Packing the sheets to a 3D structure decreases, furthermore, very
strongly the band gap value. For dicarbide sheets we identify two
candidates that could be purely metallic compounds: Sr$_3$C$_6$ and
Ba$_3$C$_6$. Both systems may have a significant amount of charge
carriers (we find this for both isolated and 3D sheet structures).
The rest of the cyanide and dicarbide compounds are found to be
either insulators or semimetals with a slight band overlap between
the occupied and unoccupied states. In the 3D packed sheet geometry,
Zn$_3$C$_6$ and Hg$_3$C$_6$ have the highest gaps ($\sim 0.7$~eV).

\begin{table}[t]
\caption{HOMO/LUMO band gap (eV) for isolated and packed cyanide and
dicarbide structures. The metallic or semimetallic character is
indicated when relevant.}
\begin{ruledtabular}
\begin{tabular}{lcccc}
Case                                        & \multicolumn{2}{c}{Chains}  & \multicolumn{2}{c}{Sheets}    \\
\cline{2-3}\cline{4-5}
                                            & Isolated    &   Packed      & Isolated    &   Packed        \\
\hline
CuCN                                        & 3.9         & 2.5           & 1.6         & 0.2             \\
AgCN                                        & 4.7         & 3.1           & 1.8         & 0.3             \\
AuCN                                        & 4.6         & 2.3           & 1.6         & 0.2             \\
\hline
BeC$_2$                                     & 3.6         & 2.9           & 0.6         & SM\footnote{Semimetallic}         \\
MgC$_2$                                     & 4.2         & 2.0           & 1.5         & SM         \\
CaC$_2$                                     & 2.6         & 1.7           & SM     & SM         \\
SrC$_2$                                     & 2.0         & 1.7           & M\footnote{Metallic}         & M             \\
BaC$_2$                                     & 1.3         & 1.6           & M         & M             \\
\hline
ZnC$_2$                                     & 3.9         & 2.4           & 1.2         & 0.7    \\
CdC$_2$                                     & 4.0         & 2.8           & 0.6         & SM \\
HgC$_2$                                     & 3.8         & 3.7
& 0.9         & 0.7
\end{tabular}
\end{ruledtabular}
\label{Table:(DOS - HOMO/LUMO gap)}
\end{table}

\subsection{Formation energy of dicarbides $M_3$C$_6$}
\label{sec:formation}

We have estimated the formation energy of the predicted new dicarbides
as the difference between their total energy and
the total energy of the corresponding pure elemental metal and graphite:
\begin{equation}
E_{\mathrm{F}} = E(M\mathrm{C}_2) - [E(M) + 2E(\mathrm{C})],
\end{equation}

\noindent where $E(M\mathrm{C}_2)$ is the total energy of the chain or
sheet structure per $M$C$_2$ unit. $E(M)$ and $E(\mathrm{C})$ are the
total energies of one atom in the zero-temperature phase of the bulk
crystal and in graphite, respectively. The total energy for graphite
was calculated to be -9.18 eV. The results are given in
Table~\ref{Table:Formation energy}, where we have also given the
atomization energies for the elemental metals, which compare
reasonably well with experimental formation enthalpies.

From a thermodynamic consideration, we see that the formation of the
dicarbides of Mg and Zn-Hg is clearly endothermic, while
CaC$_2$-BaC$_2$ are nearly thermoneutral. Of them, the {\it M}C$_2$,
{\it M} = Mg, Ca, Sr and Ba, exist. As check on the method, we
compare the calculated atomization energies for the metals. Instead
of graphite, which is a highly stable material, one could think of
different benzene-derived substances as possible starting materials
in the synthesis of the predicted new sheet-type compounds.

\begin{table}[h!]
\caption{Formation energy $E_{\mathrm{f}}$ for Group 2 and 12
sheet-type dicarbides and difference to the chains ({\it $\Delta E$}
= $E^{\mathrm{chain}} - E^{\mathrm{sheet}}$) per $M$C$_2$ unit. The
calculated atomization energy $E_{\mathrm{at}}^M$ and the
experimental atomization enthalpy $H_{\mathrm{at}}^M$ for the pure
element $M$ are also given. All values are in eV.}
\begin{ruledtabular}
\begin{tabular}{lcccccccccc}
Case                                  & $E_{\mathrm{f}}^{\mathrm{sheet}}$        & {\it $\Delta E$} & $E_{\mathrm{at}}^M$   & $H_{\mathrm{at}}^M$ (exp.$\footnote{Ref. \onlinecite{CRC-Handbook}}$) \\
\hline
Mg$_3$C$_6$                           & 1.55    &   -0.36  &  1.51 &  1.531 \\
Ca$_3$C$_6$                           & 0.61    &   -0.65  &  1.91 &  1.847 \\
Sr$_3$C$_6$                           & 0.92    &   -0.91  &  1.61 &  1.704 \\
Ba$_3$C$_6$                           & 1.05    &   -0.85  &  1.88 &  1.866 \\
\hline
Zn$_3$C$_6$                           & 1.99    &    0.67  &  1.11 &  1.355 \\
Cd$_3$C$_6$                           & 2.32    &    0.65  &  0.72 &  1.161 \\
Hg$_3$C$_6$                           & 1.99    &    1.21  &  0.08 &  0.635 \\
\end{tabular}
\end{ruledtabular}
\label{Table:Formation energy}
\end{table}

\vspace{0.7cm}
\section{Conclusions}

We have suggested a number of new solid substances, of which the
sheet-type $M_3$C$_3$N$_3$ ($M$ = Cu, Ag, Au) and the sheet-type
$M_3$C$_6$ ($M$ = Mg-Ba) may have the best chance of being prepared.
The density of the new substances is in general lower   than that of
the known chain-type structures. For the identification of these new
substances we have reported their vibrational frequencies. The
majority of the in-plane frequencies of the new compounds are
clearly separated from the stretching and bending frequencies of the
known chain-type cyanide and dicarbide structures. Evidence is found
that the Sr and Ba sheet-type compounds could be metallic both in
the 3D crystal and as isolated sheets. All the Group 2 or Group 12
sheet-type structures are endothermic. The least endothermic one is
calcium carbide. If any of these compounds can be made, the metallic
Sr and Ba ones may find the most interesting applications. The sheet
structures of lanthanides might form a further possible class of new
compounds. Note that CaC$_2$-structured, insulating or metallic {\it
Ln}C$_2$ are known for {\it Ln} = Y, La, Ce, Tb, Yb, Lu and
U.\cite{Lanthanides}

\acknowledgments

We belong to the Finnish Center of Excellence in Computational
Molecular Science (2006-2011). The grants 110571, 200903, 201291,
205967 and 206102 of The Academy of Finland are also gratefully
acknowledged. The work was also supported by the Research Funds of
the University of Helsinki.

\bibliography{pze-1_prb}

\end{document}